\documentclass[fleqn,usenatbib]{mnras}
\usepackage{comment}
\usepackage{newtxtext,newtxmath}
\usepackage[T1]{fontenc}
\DeclareRobustCommand{\VAN}[3]{#2}
\let\VANthebibliography\thebibliography
\def\thebibliography{\DeclareRobustCommand{\VAN}[3]{##3}\VANthebibliography}

\usepackage{graphicx}	
\usepackage{amsmath}	
\usepackage{caption}
\usepackage{subcaption}
\usepackage{longtable}
\usepackage{hyperref}

\title[Stellar populations of NGC~5053 using \textit{AstroSat}-UVIT]{Stellar populations of the globular cluster NGC~5053 investigated using \textit{AstroSat}-Ultra Violet Imaging Telescope}

\author[Nikitha et al.]{
K. J. Nikitha,$^{1}$,
S. Vig$^{1}$\thanks{E-mail: sarita@iist.ac.in} and S.K. Ghosh$^{2}$
\\
$^{1}$Indian Institute of Space science \& Technology, Thiruvananthapuram 695547, Kerala, INDIA\\
$^{2}$Tata Institute of Fundamental Research,  Colaba, Mumbai 400005, INDIA\\
}

\date{Accepted XXX. Received YYY; in original form ZZZ}

\pubyear{2021}

\begin{document}
\label{firstpage}
\pagerange{\pageref{firstpage}--\pageref{lastpage}}
\maketitle

\begin{abstract}

\par Globular clusters being old and densely packed serve as ideal laboratories to test stellar evolution theories. Although there is enormous literature on globular clusters in optical bands, studies in the ultraviolet (UV) regime are sparse. In this work, we study the stellar populations of a metal poor and a rather dispersed globular cluster, NGC~5053, using the UV instrument of \textit{AstroSat}, namely the Ultra Violet Imaging Telescope in three far-UV (F154W, F169M, F172M) and three near-UV (N219M, N245M, N263M) filters. Photometry was carried out on these images to construct a catalogue of UV stars, of which the cluster members were identified using \textit{Gaia} EDR3 catalogue. UV and optical CMDs help us locate known stellar populations such as BHB stars, RR-Lyrae stars, RHB stars, BSSs, SX-Phe, RGB and AGB stars. Based on their locations in the CMDs, we have identified 8 new BSS candidates, 6 probable eBSSs, and an EHB candidate. Their nature has been confirmed by fitting their spectral energy distributions with stellar atmospheric models. We believe the BSS population of this cluster is likely to have a collisional origin based on our analyses of their radial distribution and SEDs. BaSTI-IAC isochrones were generated to characterize the cluster properties, and we find that the observed brightness and colours of cluster members are best-fit with a model that is $\alpha$ enhanced with Y $ = 0.247$, $[Fe/H] = -1.9$ and age $ = 12.5\pm2.0$~Gyr. 
\end{abstract}




\section{Introduction}

\par Globular clusters (GCs) are old stellar systems containing millions of stars, with ages > 10 Gyr. They serve as quintessential laboratories to test stellar evolution theories \citep[e.g., see][]{gratton2019}. For long it was thought that GCs hosted a Simple Stellar Population (SSP), i.e., an ensemble of stars with the same age and chemical composition. But with recent photometric and spectroscopic results we now know that GCs host Multiple Stellar Populations \citep[MSPs;][ and references therein]{2004grattonsndencaretta,2012grattoncarettabragaglia,2016charbonnel}.Various theories have been proposed for the formation mechanisms of GCs to explain these observations. But these have not been able to account for all the observed spectroscopic and photometric characteristics of these systems \citep{2018bastian}.

\par Globular clusters have been studied extensively in the optical wavebands, but measurements in ultraviolet (UV) have been relatively sparse. The advantage of performing UV studies on a globular cluster is that we can explore certain evolved phases of stellar evolution. The primary contributors to the UV flux from a GC are Horizontal Branch (HB) stars, Blue Straggler Stars (BSSs), Post-early Asymptotic Giant Branch (PeAGB) stars, and White Dwarfs (WDs) \citep{zinn1972}. Since these stellar populations are brighter at shorter wavelengths owing to their higher effective temperatures ($\sim 5000-40,000$~K), these are usually called ‘hot stars’ \citep{moehler2001}. 

\par Horizontal branch stars are core He burning low mass stars with a hydrogen rich envelope surrounding the core. The HB is comprised of subpopulations separated by gaps in the colour-magnitude diagram (CMD): red horizontal branch (RHB) stars, blue horizontal branch (BHB) stars, and extreme horizontal branch (EHB) stars \citep{buo1986,1995castellani}. The RHB and BHB stars lie on either side of the RR~Lyrae instability strip with the RHB stars occupying the cooler side. The effective temperature (T$_{eff}$) for RHB and BHB stars range between $\sim 5000-6000$~K \citep{sneden2010} and  $\sim8000-20,000$~K \citep{heber1997}, respectively. EHB stars have a central He burning core with an envelope that is very thin to sustain hydrogen burning. They have $T_{eff} > 20,000$~K and lie at the extreme hot end of the HB \citep{ostensen2009}. After He core exhaustion, depending on the mass of the stellar envelope, the HB stars may evolve into: (i) AGB phase - where the stars undergo thermal pulsation and eventually evolve to higher temperatures with constant high luminosity, (ii) PeAGB phase - where the stars do not undergo thermal pulsation and evolve towards higher temperatures at constant but lower luminosities, or (iii) AGB-manqu\'e phase - where they do not ascend the AGB and evolve directly towards the WD cooling curve \citep{1970iben,Greggio1990,brocato1990}. The Main Sequence (MS) and Red Giant Branch (RGB) stars that constitute the majority of stars in GCs are relatively faint at UV wavelengths and this eventually results in the UV image of a GC being less crowded than the corresponding optical image. 

\par Space based UV missions are essential for measurements at shorter wavelength ranges ($\lambda<300$~nm) due to the effect of atmospheric absorption. The Orbiting Astronomical Observatory-2 \citep[OAO-2;][]{1980welch} and the Astronomical Netherlands Satellite \citep[ANS;][]{1981van} were among the early UV missions to study the GCs. Some of the other UV space missions include the International Ultraviolet Explorer \citep[IUE; e.g.][]{1982IUE, 1993altner}, the Ultraviolet Imaging Telescope \citep[UIT; e.g.][]{1992hill, 1997dorman, 1998parise}, SWIFT - Ultraviolet/Optical Telescope \citep[UVOT; e.g.][]{2007holland} and \textit{Galaxy Evolution Explorer} \citep[GALEX;][]{schiavon2012}. The Space Telescope Imaging Spectrograph (STIS) and the Wide Field Camera 3 (WFC3) aboard the Hubble Space Telescope (HST) have been able to analyze the core regions of certain GCs extensively in the far-UV (FUV) wavelengths \citep[e.g.][]{2002knigge,2017dieball} thereby revolutionizing our understanding of the UV bright stellar populations. The latest addition to this group of space based UV missions is the \textit{AstroSat} mission with its Ultra Violet Imaging Telescope (UVIT) having a field-of-view of $\sim28'$ and a spatial resolution of $1.8''$. In recent years a number of GCs have been studied using the UVIT thereby proving its capability in surveying the low density GCs as well as the outer areas of more centrally concentrated GCs \citep[e.g.][]{2016subra_astro,2017subra_astro,2019sahu_astro,2019jain_astro}. 

\par The UV studies of GCs, although relatively few, provide significant information regarding the constituent hot stellar populations. The FUV CMDs of HB stars display peculiarities (eg. gaps) which were investigated by \citet{1998ferraro}. Studies by \citet{2011dalessandro,2015lagioia,2016Abrown} have demonstrated how the FUV CMDs are the best suited to examine the temperature dependency of the HB morphology in GCs as compared to the optical CMDs. UV CMDs also provide crucial information regarding exotic stellar populations like the EHB stars and blue-hook stars \citep[e.g.][]{1998whitney,2009dieball,2010brown,2012brown}. UV observations have also proved to be essential tools to investigate the BSSs. These stars are known to occupy a region blueward of the main sequence turn-off (MSTO) point in the CMD defying the canonical stellar evolution within a cluster. Known BSS stars were not only identified in the UV bands, new BSS candidates were also detected for the first time using FUV \citep[e.g.][]{2001ferraro,2002knigge,2005dieball,2010dieball}. These authors have also analysed the radial distribution of BSSs, thereby attempting to understand their formation mechanisms. The hypotheses for BSS formation mainly include a mass-transfer mechanism or a collisional mechanism. FUV spectroscopic study on 47 Tuc by \citet{2008knigge} showed the first ever BSS-WD binary system to be known in any GC. Identification of such hot companions to BSSs have proved to be crucial in understanding how they formed, especially in binary systems \citep[e.g.][]{2014gosnell,2015gosnell}.

\par In the current study, we examine the stellar populations of NGC~5053 using the UVIT, on-board the \textit{AstroSat}. NGC~5053 is a galactic globular cluster that lies in the northern constellation of Coma Berenices, having Galactic coordinates $l=335.70^\circ$, $b=78.95^\circ$. This cluster is located at a distance of $17.54\pm0.23$~kpc \citep{2021baumgardt} and the metallicity is estimated to be $[Fe/H]=-2.27$ \citep{harris2010}. It is, therefore, placed amongst the metal-poor galactic globular clusters. The tidal radius of this cluster has been estimated as $r_t = 15.2 \pm 3.3'$ by \citet{2019boer}. When it was discovered in 1784, it wasn't classified as a globular cluster because of its appearance and structure; there was no densely packed nucleus, and the central region was resolvable \citep{1786herschel}. These observations placed this GC in a grey-area between globular and open clusters. It was only in 1928 that Baade first classified it as a GC owing to its high latitude, richness in faint stars and presence of variable stars \citep{baade1928}. The variable stars in this cluster include RR~Lyrae variables and SX~Phoenicis (SX~Phe) stars \citep{sawyer1946,nemec1995,nem1995,nemec2004}. The cluster is known to have a HB that is extended predominantly towards the blue flank of the RR~Lyrae instability strip, which is an archetypal feature in many metal poor globular clusters \citep{sara1995}. A rich population of BSSs has been identified in this cluster \citep{sara1995}. The presence of relatively large number of BSSs in such low density GCs has been examined to propose alternate formation mechanisms of BSSs \citep{1991leofahl}. The cluster has been examined in the UV regime by \citet{schiavon2012} using photometric data from GALEX, and they have published a catalogue of the UV bright stars present in this cluster. The location of this cluster in the sky and its velocity have given rise to a debate on whether it belongs to the Milky Way or Sgr~dSph \citep[e.g.][]{law2010,boberg2015,Sbordone_2015,baitian2018}.

\par The organization of the paper is as follows. In Sect.~\ref{sec:obs}, the UV observations and data reduction procedures are outlined. Sect.~\ref{sec:ident} describes the identification of the stellar populations. The results from observations and modelling are elucidated in Sect.~\ref{sec:resobs} and \ref{sec:resmodel}, respectively. The results are analysed in Sect.~\ref{sec:disc}, and a brief summary is presented in Sect.~\ref{sec:conc}. Supplementary table and figures are included in the Appendix.

\section{Observations and Data Reduction}
\label{sec:obs}
\subsection{\textit{AstroSat} - UVIT Observations and data reduction}

\par The images of NGC~5053 were obtained using the Ultra Violet Imaging Telescope (UVIT) instrument on-board \textit{AstroSat}. \textit{AstroSat} is India's first space based multi-wavelength mission which is capable of observing simultaneously in UV, visible and X-rays \citep{2agrawal2004}. It has five payloads including UVIT which operates in ultraviolet wavelength range. It is primarily an imaging instrument and consists of a twin telescope system in the Ritchey-Chretien configuration. One telescope observes in the far-ultraviolet (FUV; $130-180$~nm), while the other telescope observes in the near-ultraviolet (NUV; $200-300$~nm) and the visible (VIS; $320-550$~nm), the latter is made possible using a dichroic beam splitter.  The field-of-view is $28'$, and the spatial resolution is $<1.8''$ for the UV bands, $\sim2.2''$ for the VIS band. Within each of the three bands, a number of filters with smaller passbands are available. A detailed description of the UVIT instrument can be found in \citet{Tandon_et_al.2017b}.
\begin{figure*}
         \includegraphics[width=\textwidth]{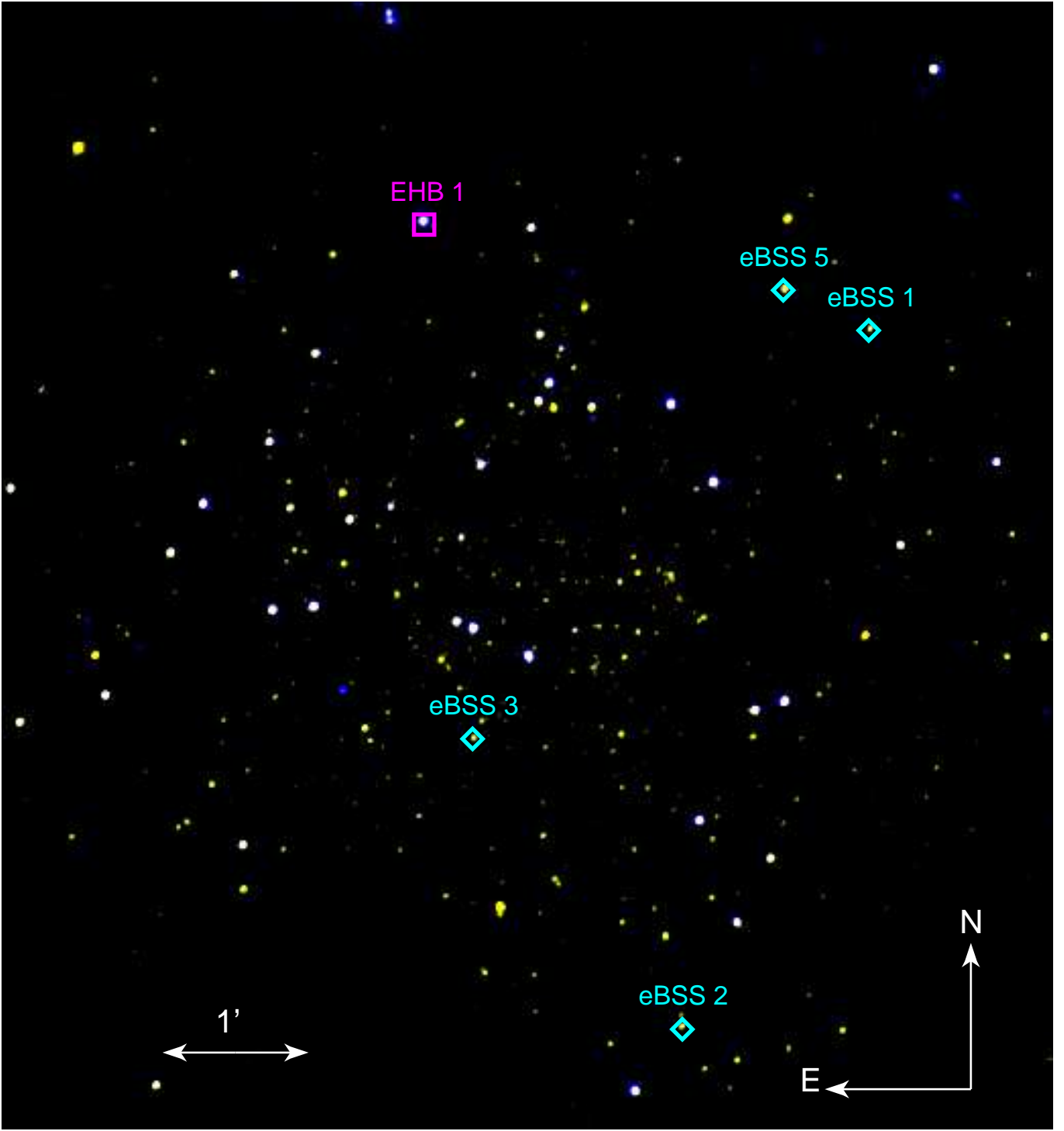}
         \caption{The UVIT image of NGC~5053. The blue and yellow colours denote emission in F154W and N263M filters, respectively. The location of the EHB candidate and a few eBSS candidates are marked in the figure.}
         \label{fig:fig1}
\end{figure*}
\par The UVIT data of the GC presented in this work are of observations carried out in two epochs. The imaging of the GC through the FUV filters: F154W and F172M, and the NUV filters: N219M, N245M, N263M were taken on June 14, 2017. Observations through the FUV filter: F169M were obtained on February 13, 2020. The details of the observations are given in Table~\ref{tab:obssum}. The Level-1 (L1) data which includes raw exposures from individual channels of UVIT, were made available by the Indian Space Science Data Center (ISSDC/ISRO). This L1 data was processed using the UVIT Level-2 Pipeline (version UL2Pv6.3) to obtain sky image products for each of the filters in NUV and FUV bands used in the present study. The Level-2 pipeline corrects for the spacecraft drift as well as various instrumental systematic effects within the UVIT \citep{ghosh2021}. While this pipeline at its final processing block attempts astrometric correction using brighter UV stars detected in the images, it was not successful for the images in the FUV filters. Accordingly, we applied an alternate scheme for astrometric corrections.
The images in all the six filters were corrected for astrometry by comparing them with the USNO~A2.0 catalogue \citep{monet1998}. The {\tt CCMAP} task in IRAF was executed to perform astrometry. The uncertainties in positions, post-astrometry, were within $1''$. The UVIT image of NGC~5053 created using N263M and F154W filters is displayed in Fig.~\ref{fig:fig1}.
\begin{table*}
\centering
\caption{Details of UVIT filters and photometry of sources extracted from the images. }
\label{tab:obssum}
\begin{tabular}{lcccccccc}
\hline
Band & Filter & Alternate & $\lambda_{mean}$ &$\delta\lambda $ & ZP mag & Exposure Time & Sensitivity Limit & Number of\\
& & Name & (\AA) & (\AA)& & (sec) & (AB mag for SNR=3) & stars\\
\hline
FUV & F154W & BaF$_2$ & 1541 & 380 & 17.765$\pm$0.01 & 1157 & 24.05 & 202\\
FUV & F169M & Sapphire & 1608 & 290 & 17.453$\pm$0.01 & 1566 &  24.22 & 252\\
FUV & F172M & Silica & 1717 & 125 & 16.341$\pm$ 0.02 & 509& 22.58 & 93\\
NUV & N219M & NUVB15 & 2196 & 270 & 16.59$\pm$0.02 & 1167 & 22.83 & 727\\
NUV & N245M & NUVB13 & 2447 & 280 & 18.50$\pm$0.07 & 583 & 23.87 & 1215\\
NUV & N263M & NUVB4 & 2632 & 275 & 18.18$\pm$0.01 & 843 & 23.75 & 1392\\
\hline
\end{tabular}
\end{table*}

\subsection{Photometry}
We performed aperture photometry on the UV images using two softwares: {\tt  DAOPHOT} \citep[DAO;][]{1987stetson} and {\tt Source Extractor} \citep[SE;][]{1996bertin}. The selection of stars and photometry by both methods were consistent for most of the stars. Towards the fainter end, the detection by SE was relatively more efficient than the {\tt DAOPHOT}, hence SE photometry was carried out for the fainter stars. The AB magnitudes of stars were estimated using the formula:
\begin{equation}
    m_{AB}=m_{ZP} - 2.5*log(CPS)
\end{equation}
where m$_{ZP}$ is the zero-point magnitude of the filter taken from \citet{Tandon_et_al.2017b} and CPS is the integrated counts/second for a star through the filter. In addition to these two methods, we visually inspected the field through each filter to locate faint sources that were not detected by both the softwares. The aperture photometry of these sources were carried out subsequently. 

\par The errors in AB magnitudes were estimated as a combination of the error in the zero-point magnitude and the error in the flux estimation; for the latter, the uncertainties in the counts are dominated by shot noise of the photon-counting detector. The ZP magnitudes as well as their uncertainties are displayed in Table \ref{tab:obssum}.

\par The AB magnitudes were corrected for reddening by applying the extinction law of \citet{Cardelli1989}. Adopting the ratio of total-to-selective extinction as R$_v= 3.1$ for the Milky Way \citep{whitford1958} and the value of foreground reddening E$(B-V) = 0.0149$ from the extinction map of \citet{schlegel1998}, we obtained the extinction coefficient in the visible as A$_V =  0.05$. Using this, the extinction coefficients in different filters, A$_{\lambda}$, were estimated and the magnitudes were dereddened. 
 Finally, we compiled the stars that were detected through each filter to construct a catalogue of stars detected in UVIT images. In the current study, we consider those stars that are brighter than the magnitude corresponding to a signal-to-noise ratio (SNR) of 3 to have a magnitude limited sample through each filter. The AB magnitudes corresponding to this sensitivity limit are listed in Table~\ref{tab:obssum}. These values have not been corrected for reddening. The total number of UV detected stars in the catalogue is 1871. The number of stars detected through each filter is also listed in Table~\ref{tab:obssum}. 

\section{Results from observations}
\label{sec:resobs}

\subsection{Identification of stellar populations}
\label{sec:ident}
\par We have identified the stellar populations in our UVIT catalogue from various photometric and spectroscopic studies available in literature. We identified HB stars, BSSs and giant stars based on the BVI CCD photometry by \citet{sara1995}. The RR~Lyrae variables and SX~Phe variable stars belonging to this cluster were compiled from the catalogue of all known variable stars \citep{samus2017}. Spectroscopic studies by \citet{boberg2015} and \citet{kirby2016,kirby2018} were used to identify the giant stars in the cluster that were primarily detected in the longer wavelength NUV filters. \citet{schiavon2012} used UV CMDs of GALEX to present a catalogue of post-core He burning stellar candidates, which include post-AGB (PAGB) stars, PeAGB stars, and AGB-manqu\'e stars of 44 GCs. Towards this cluster, no candidates were found by them. Altogether we were able to identify 35~BHB stars, 8~RHB stars, 10~RR~Lyrae stars, 16~BSSs, 4~SX~Phe stars, 5~AGB stars and 59~RGB stars in the UVIT images. The UVIT photometric data of these stellar populations are available in electronic form. An example of this is shown in Appendix (Table~\ref{tab:photo_catalogue}).

\subsection{Optical Colour-Magnitude Diagram using \textit{Gaia} counterparts}
\par We searched for optical counterparts to the stars detected in UVIT images from the \textit{Gaia} Early Data Release 3 (EDR3) catalogue \citep{2016gaia,2021gaia}. We used a search radius of  $1''$ and considered the closest object as an association in cases where two objects were found within the search radius. We found a total of 909 \textit{Gaia} counterparts. We observe that among the 909 sources, only 42 sources have a second closest counterpart within $1''$ and the minimum separation of this second closest counterpart is $\sim 0.5''$ observed in only 4 sources. We therefore believe that the chance of misidentification is quite low. This is also evident from the optical as well as the UV CMDs where the identified populations occupy the expected loci, as discussed in later sections. We plotted these stars in the optical CMD using the two \textit{Gaia} passbands: BP and RP. The BP or Blue Photometer passband is a narrow passband covering a wavelength region between $330-680$~nm, while the RP or Red Photometer passband is a passband covering the wavelength region between $640-1000$~nm. This optical CMD is likely to be contaminated with stars that don't belong to the cluster. In order to identify the members of the cluster, we carried out a proper motion (PM) analysis of the stars.

\par
\begin{figure}
     \centering
         \includegraphics[width=\columnwidth]{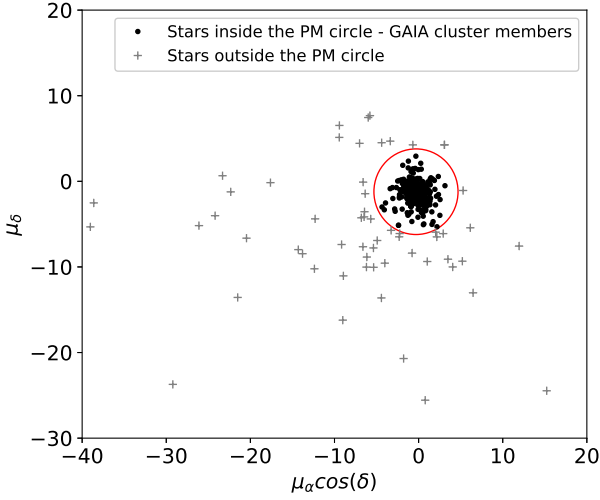}
         \caption{The Vector Point Diagram (VPD) for the UVIT-\textit{Gaia} sources. The region within the circle of radius 5~mas/yr centered around the mean PM of the cluster is used to select the cluster members.}
         \label{fig:fig2}
\end{figure}

\begin{figure}
         \centering
         \includegraphics[width=\columnwidth]{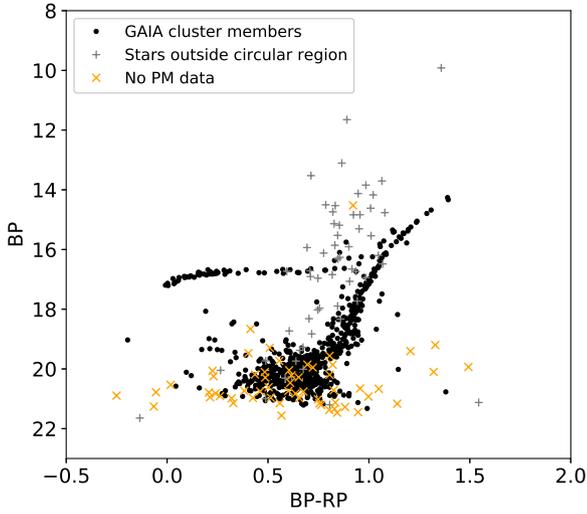}
         \caption{The \textit{Gaia} optical CMD showing different populations. The black dots represent cluster members  within the 5~mas/yr radius circle in the proper motion vector point diagram (VPD). The grey '+' sign represents stars that lie outside this circle in the VPD. The yellow crosses represent stars which are cross identified from the \textit{Gaia} catalogue but do not have PM data.}
         \label{fig:fig3}
\end{figure}

 We could extract PM data for 826 stars out of 909 stars. We plot a vector point diagram (VPD) of these stars with PM along the right ascension ($\mu_{\alpha ^*}$= $\mu_\alpha$ cos$\delta$) versus the PM along the declination ($\mu_\delta$). This is shown in Fig.~\ref{fig:fig2}. The mean PM of the cluster as determined from \textit{Gaia} is ($\mu_{\alpha}$, $\mu_\delta$) = (-0.329, -1.214) mas/yr \citep{vasiliev2021gaia}, and we find that the stars in the \textit{Gaia} VPD are clustered about this mean. We consider stars within a circle of radius 5~mas/yr about the mean as probable cluster members. The optical CMD with the cluster members selected using the PM analysis is shown in Fig.~\ref{fig:fig3}.

\par
\begin{figure}
        \centering
        \includegraphics[width=\columnwidth]{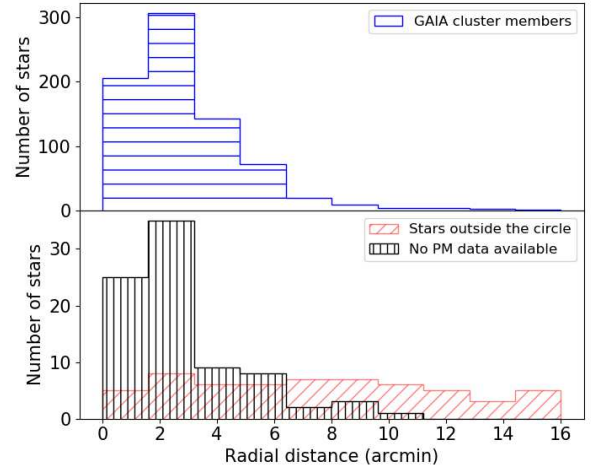}
        \caption{The radial distribution of the UVIT-\textit{Gaia} cross matched stars.}
        \label{fig:fig4}
\end{figure}

 In order to understand the membership of \textit{Gaia} counterparts other than the cluster members, we compare the radial distribution of various groups of stars: (i) the cluster members, (ii) stars with PM that lie outside the circle in the VPD, and (iii) stars for which the PM data are not available. This comparison is shown in Fig.~\ref{fig:fig4}. We note that the radial distribution of stars lying outside the circle is flat suggesting that these stars are likely to be field stars. The distribution of \textit{Gaia} stars for which PM data is not available appears consistent with the radial distribution of cluster members. We, therefore, believe that a major fraction of these stars are cluster members. 


\par
\begin{figure*}
         \centering
         \includegraphics[width=\textwidth]{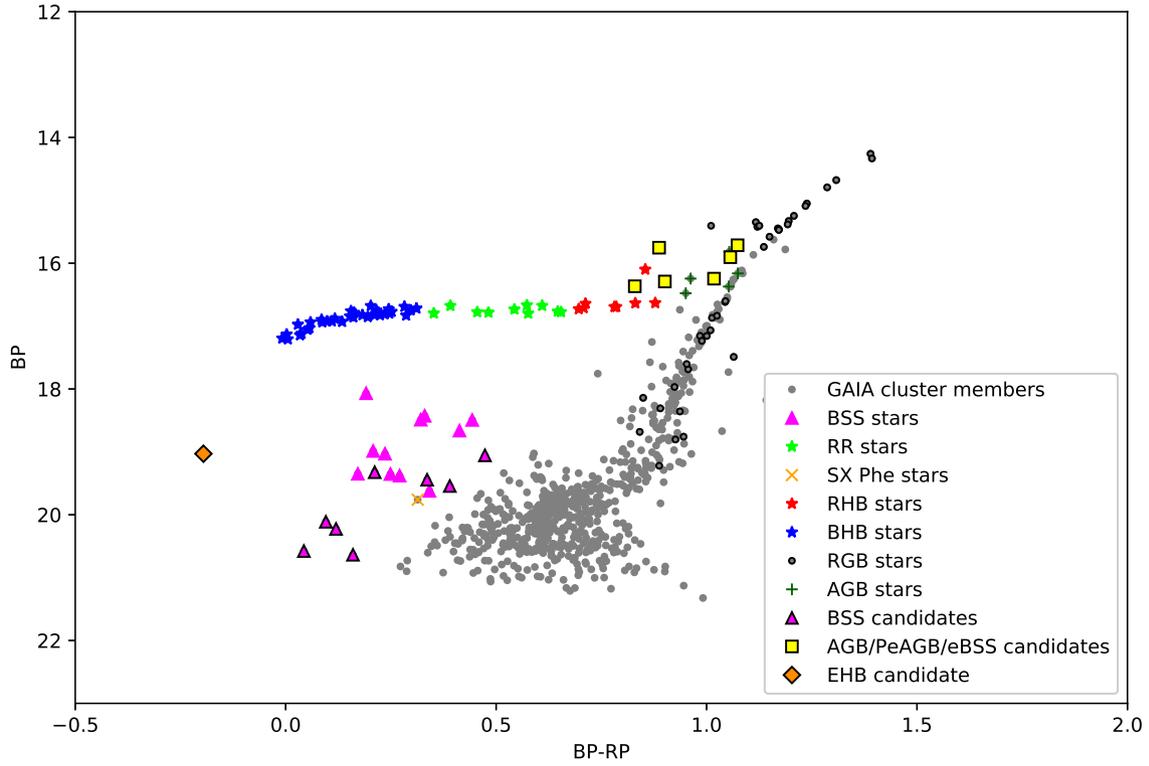}
         \caption{The optical \textit{Gaia} CMD constructed using the cluster members. The BSS candidates are marked as magenta triangles with a black outline. The AGB/PeAGB/eBSS candidates are shown as yellow squares. The EHB candidate at the extreme end of the HB is marked in orange.}
         \label{fig:fig5}
\end{figure*}

 We first discuss the optical \textit{Gaia} CMD of the cluster members detected in the UV filters. We construct a clean CMD using only the cluster members, and this is displayed in Fig.~\ref{fig:fig5}. The stars in various phases of evolution that have been identified from literature in Sect.~\ref{sec:ident} are shown in different colours. We observe an extended and well-populated horizontal branch in the CMD. The RHB, RR~Lyrae and BHB stars fully sample the branch in the BP-RP colour range of $0.0 - 0.9$~mag. The BP magnitude of the horizontal branch is $\sim16.7$~mag. The RGB branch extends from 19.2 to 14.3 mag in the BP filter. We observe the AGB stars are grouped at the intersection of the HB and RGB. Below the HB, we observe the BSSs in the BP-RP colour range of $0.2 - 0.5$~mag. 

\par There are a number of stars that have not been identified as belonging to any particular stellar population from our literature survey, and we discuss them here. There is a group of unidentified stars in the region defined by $19.1 \lesssim \textrm{BP} \lesssim 20.6$ and $0.0 \lesssim \textrm{BP-RP} \lesssim 0.5$. Based on their proximity to BSSs in the CMD, these stars are potential BSS candidates. We also note a group of unidentified stars lying in the region $15.7 \lesssim \textrm{BP} \lesssim 16.4$ and $0.8 \lesssim \textrm{BP-RP} \lesssim 1.1$. Their location in the CMD suggests that they are likely to be AGB, PeAGB or evolved-BSS (eBSS) candidates. In addition, we observe a star that lies at the extreme hot end of the horizontal branch, at $\textrm{BP}=19.0$ and $\textrm{BP-RP}=-0.2$. We believe this star to be a potential EHB candidate based on the observed morphologies of HBs in optical CMDs \citep[e.g.][]{1996dcruz,Moehler_2004}. These stars are further analyzed using the UV CMDs in later sections.
\begin{figure*}
\centering
\begin{subfigure}{0.5\textwidth}
    \includegraphics[width=\textwidth]{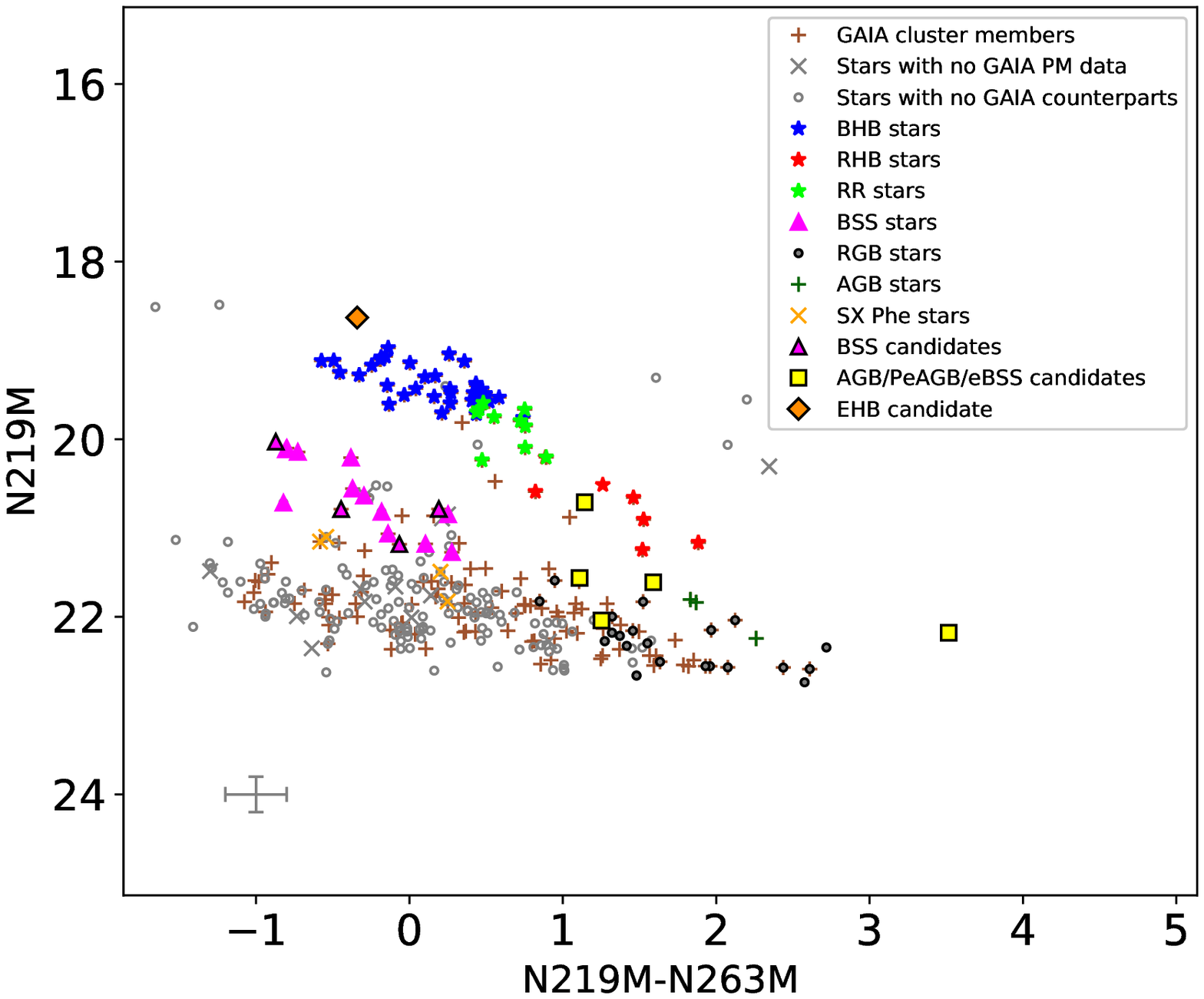}
    \caption{NUV-NUV CMD.}
    \label{fig:6a}
\end{subfigure}
\hspace{-0.2cm}
\begin{subfigure}{0.5\textwidth}
    \includegraphics[width=\textwidth]{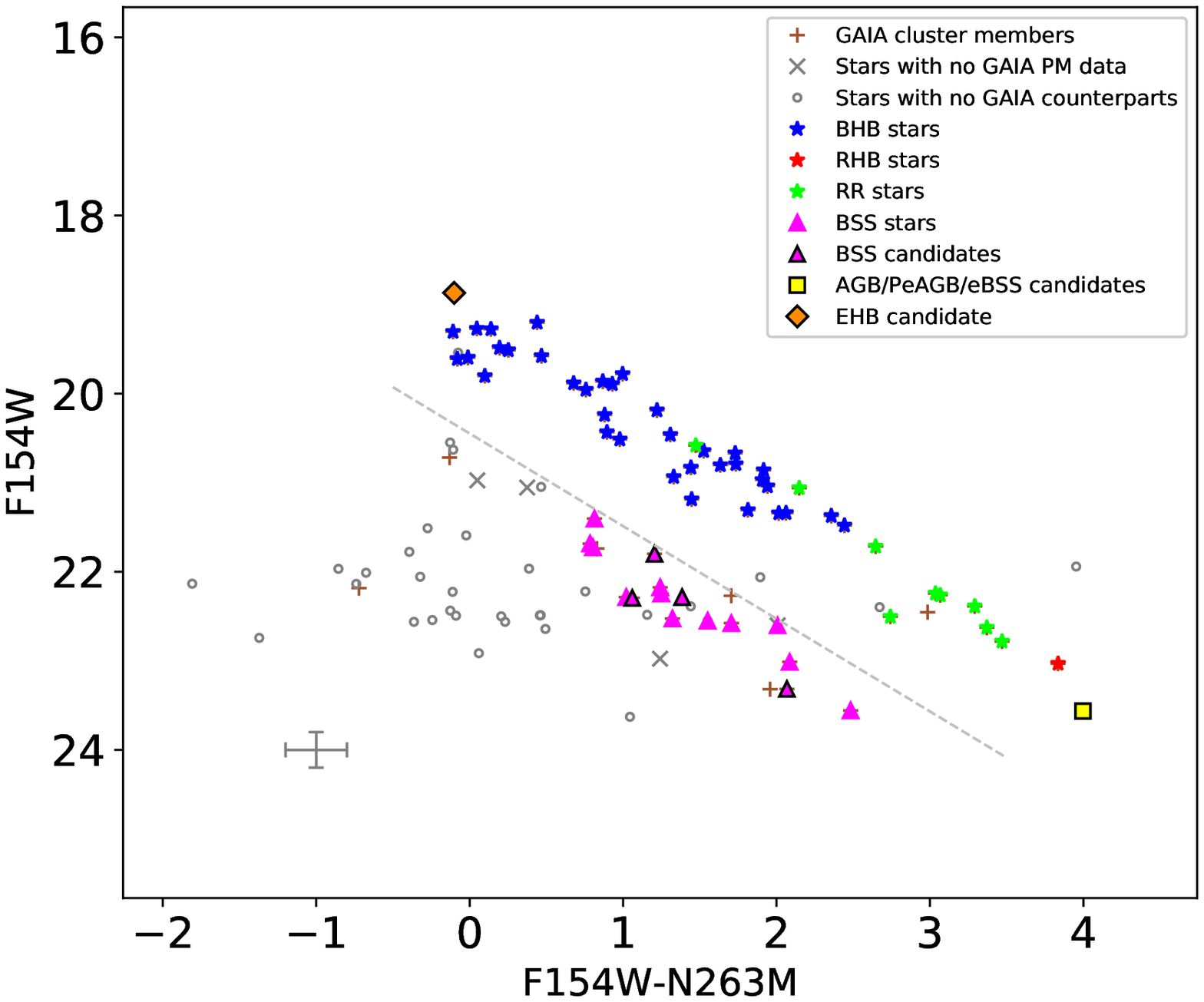}
    \caption{FUV-NUV CMD.}
    \label{fig:6b}
\end{subfigure}
\hspace{-0.2cm}
\caption{The UVIT NUV-NUV CMD constructed using N219M and N263M filters (left) and the FUV-NUV CMD constructed using F154W and N263M filters (right). The typical error along the magnitude and colour axis is shown in the bottom left corner of the CMDs.}
\label{fig:uvcmd}
\end{figure*}

\subsection{Ultraviolet Colour-Magnitude Diagrams}

\par In this section, we analyse stellar populations belonging to this cluster in the UV CMDs. As we have three NUV and three FUV filters, a large number of CMDs are possible. Here we discuss two CMDs in which the stellar populations can be discerned quite distinctly. 
\begin{table*}
\centering
\caption{Details of EHB/BSS/eBSS candidates detected in NGC~5053 in this work are listed below. Their coordinates and photometry in UVIT and \textit{Gaia} filters are tabulated. The last column indicates the positional difference between UVIT and \textit{Gaia} of the sources. It is to be noted that the detection of a source in a filter image depends on the exposure time as well as the effective area of the filter. }
\label{tab:candidates}
\begin{tabular}{lccccccccccc}
\hline
Source & RA & DEC & F154W & F169M & F172M & N219M & N245M & N263M & BP & RP & $\Delta$\\
 & (deg) & (deg) & (mag) & (mag) & (mag) & (mag) & (mag) & (mag) & (mag) & (mag) & (arcsec)\\
\hline
EHB 1 & 199.1259	& 17.7438 & 18.99 & 19.23 & 20.11 & 18.92	& 19.23 & 19.28 & 19.03 & 19.23 & 0.3\\
BSSc 1 & 198.9949 & 17.5896 & 20.93	& 22.23 & 21.45 & 21.37 & 20.75 & 20.69 & 19.05 & 18.58 & 0.8\\
BSSc 2 & 199.1680 & 17.6271 & 20.94 & - & -& 21.41&-& 21.33 & 19.32 & 19.11 & 0.0\\
BSSc 3 & 199.1081 & 17.9393 & 20.18 & 21.71 &	- & 20.41 & 20.05 & 20.99 & 19.44 & 19.11 & 0.5\\
BSSc 4 & 199.1080 & 17.6864 & - & - & - & - & 22.28 & 21.17 & 19.54 & 19.15 & 0.5\\
BSSc 5 & 199.1470 & 17.6876 & - & 23.48 & - & - & - & 21.43 & 20.11 & 19.16 & 0.8\\
BSSc 6 & 199.0700 & 17.6824 & 21.33 & - & 21.49 & 21.82 & 22.36 & 21.34 & 20.23 & 20.11 & 0.8\\
BSSc 7 & 199.0280 & 17.6931 & - & 22.08 & 21.32 & - & - & 21.97 & 20.64 & 20.49 & 0.7\\
BSSc 8 & 199.1576 & 17.6713 & - & 22.61 & 21.54 & - & 23.01 & 22.10 & 20.58 & 20.53 & 0.8\\
eBSS 1 & 199.0724 & 17.7315 & - & 22.40 & - & 21.76 & 21.48 & 20.12 & 16.24 & 15.23 & 0.4\\
eBSS 2 & 199.0951 & 17.6521 & - & 22.78 & - & 22.32 & 21.31 & 18.76 & 15.90 & 14.85 & 0.3\\
eBSS 3 & 199.1199 & 17.6850 & - & 22.14 & - & - & 21.46 & 19.90 & 15.71 & 14.64 & 0.4\\
eBSS 4 & 199.3318 & 17.7784 & - & 22.50 & - & 21.44 & 20.90 & 20.20 & 16.34 & 15.54 & 0.9\\
eBSS 5 & 199.0827 & 17.7361 & 23.69 & 22.08 & 24.98 & 21.55 & 21.00 & 20.25 & 15.75 & 14.86 & 0.3\\
eBSS 6 & 199.2250 & 17.6263 & - & 22.95 & - & 21.46 & 20.96 & 20.28 & 16.29 & 15.39 & 0.4\\
\hline
\end{tabular}
\end{table*}
\par The NUV-NUV CMD constructed using N219M and N263M filters is presented in Fig.~\ref{fig:6a}. The different populations identified earlier in Sect.~\ref{sec:ident} appear quite segregated in this CMD. The HB population consisting of  BHB, RR~Lyrae and RHB form a well defined diagonal sequence in the CMD lying in the region defined by $18\lesssim \textrm{N219M} \lesssim 22$ and $-1 \lesssim \textrm{N219M-N263M} \lesssim 2$. The BSS population along with the associated variable stars, SX~Phe, lie just below this sequence in the region defined by $20 \lesssim \textrm{N219M} \lesssim 22$ and $ -1 \lesssim \textrm{N219M-N63M} \lesssim 0.5$. The potential EHB candidate lies at the extreme hot end of the HB sequence in this NUV-NUV CMD, at $\textrm{N219M}=18.6$ and $\textrm{N219M-N63M}=-0.3$,  according to expectations. We observe the BSS candidate population to be concentrated within the region defined by the BSSs thereby confirming them as BSS candidates. The PAGB/PeAGB/eBSS candidates in the NUV-NUV CMD are fainter than the HB although there is some overlap with RHB stars. In general, they are scattered in the region occupied by the giants. A similar trend was observed for other NUV-NUV CMDs as well. 

\par The FUV-NUV CMD constructed using the F154W and N263M filters is displayed in Fig.~\ref{fig:6b}. The HB stars lie in the region defined by $ 19\lesssim \textrm{F145W} \lesssim 23$ and $-0.2 \lesssim \textrm{F154W-N263M} \lesssim 4$. The EHB candidate lies at the the hottest end of the HB in the FUV-NUV CMD, at $\textrm{F154W}=18.9$ and $\textrm{F154W-N263M}=-0.1$, thereby providing strong support to our claim of this object being an EHB star. The BSS population occupies the region defined by $21 \lesssim \textrm{F154W} \lesssim 24$ and $0.7 \lesssim \textrm{F154W-N263M} \lesssim 2.5$. From the CMD, we note that the HB and BSS populations are clearly segregated as two parallel sequences, separated by a clear gap of about $\sim0.6$~mag in F154W-N263M colour throughout. This gap can be described by a straight line (shown as solid grey line in Fig.~\ref{fig:6b}) with an equation given by:
\begin{equation}
F154W = 1.0 * (F154W-N263M) + 20.1
\end{equation} 
The BSS candidates emitting in FUV lie in the region occupied by BSSs in the FUV-NUV CMD, thus reinforcing our assertion. A star belonging to the AGB/PeAGB/eBSS candidate group is found to lie at the fainter and cooler end of the HB sequence. We show in later sections that it is highly likely to belong to eBSS stars. The giant stars are not detected in the FUV filters. The coordinates along with their optical and UV magnitudes of these candidates are given in Table~\ref{tab:candidates}.

\section{Results from modelling}
In this section we outline the tools and methods employed to describe the properties of the cluster and the physical parameters of the UV bright populations of the cluster.

\label{sec:resmodel}
\subsection{Characterizing the properties of the cluster}

\par A common tool used to characterize the cluster properties is the fitting of isochrones to the cluster CMD \citep[e.g.][]{2018Pbarker,2019gontcharov}. In the current study we have used Bag of Stellar Tracks and Isochrones (BaSTI-IAC) models\footnote{\url{http://basti-iac.oa-abruzzo.inaf.it/index.html}}, to diagnose the cluster properties. BaSTI is a library of stellar tracks and isochrones, that has been extensively used to study the star clusters, field stars and galaxies \citep{Hidalgo_2018,Pietrinferni_2021}. It includes models for solar scaled heavy element mixture and $\alpha$ enhanced heavy element mixture. We generated BaSTI isochrones corresponding to the \textit{Gaia} CMD. NGC~5053 being a metal poor cluster will have stars with significant enhancement of $\alpha$ elements. This is also consistent with results from spectroscopic studies \citep[eg., see][]{mash2019}. Consequently, we have used the $\alpha$ enhanced heavy element mixture models. The only $\alpha$ enhanced heavy element mixture model that is available is that for which $[\alpha/Fe] = +0.4$. For this heavy element mixture, the available grids are models that include the processes of overshooting, diffusion and mass loss characterized by mass loss parameter of $\eta=0.3$ with varying He mass-fractions corresponding to Y = 0.247, 0.275, and 0.300. For each He mass-fraction model, there is a subset of models with specific metallicities and age ranges. 
 \begin{figure}
    \centering
    \includegraphics[width=\columnwidth]{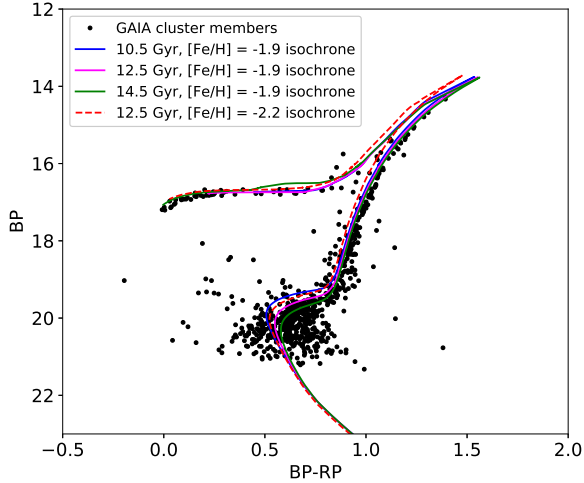}
    \caption{The optical \textit{Gaia} CMD showing the cluster members. Overlaid are BaSTI isochrones computed for ages 10.5~Gyr, 12.5~Gyr and 14.5~Gyr for a metallicity of $[Fe/H] =-1.9$. The red dashed line represents the isochrone corresponding to a metallicity of $[Fe/H] =-2.2$ for age of 12.5~Gyr.}
    \label{fig:fig7}
\end{figure}
\par The input parameters to the models include distance modulus, age and metallicity. From the literature, we find a range of values of these parameters \citep{bell1983,Zinn1985,suntzeff1988,armandroff1992,sara1995,geisler1995,rutledge1997, Rosenberg1999,Salaris2002,nemec2004,Ferro2010,Sbordone_2015,boberg2015}. The range of distance modulus values lie between 15.8 and 16.2, the metallicities inferred are between $-2.8$~dex and $-1.8$~dex, while the age of the cluster is estimated to be between 10 and 14.5~Gyr. Based on these values, we constructed isochrones of models with Y = 0.247, 0.275, 0.300, [Fe/H] = -2.5, -2.2, -1.9 dex, age range $10 - 14.5$~Gyr and distance modulus values of 16.08, 16.12, 16.16. Amongst these models, we find the isochrone model that fits the observed CMD reasonably well has He fraction = 0.247, $[Fe/H] = -1.9$~dex, distance modulus = 16.16 with age of $12.5 \pm 2.0$~Gyr. The BaSTI isochrones corresponding to ages 10.5, 12.5 and 14.5~Gyr for the above mentioned parameters are presented in Fig.~\ref{fig:fig7}. We have also plotted the isochrone corresponding to $[Fe/H] = -2.2$~dex for age of 12.5~Gyr for comparison. We can see that the latter does not fit the giant branch and the MSTO, and is a poorer fit in comparison to the model with $[Fe/H] = -1.9$~dex.

\subsection{Characterizing the properties of stellar populations}\label{sec:stellarchar}

\par We attempt to understand the nature and properties of the stellar populations of the cluster by analysing their photometry across multiple bands. In other words, we construct the spectral energy distributions (SED) of HB stars, BSSs, their respective candidates, and eBSS candidate, and fit them with stellar atmospheric models to extract their likely parameters. The SEDs in the current study were constructed using UV and optical photometric data from the UVIT catalogue, \textit{Gaia} EDR3 catalogue~\citep{2016gaia,2021gaia}, Panoramic Survey Telescope and Rapid Response System (Pan-STARRS) survey-DR1 catalogue \citep{2016chambers} and GALEX-DR5 catalogue \citep{2011bianchi}, if available. Similar to \textit{Gaia}, we employed a search radius of $1''$ to locate the stellar counterparts in Pan-STARRS and GALEX catalogues.

\par We utilize the Virtual Observatory SED Analyzer (VOSA) to perform the SED fitting with the theoretical models. VOSA \citep{Bayo2008} generates synthetic photometry from theoretical spectra and compares it with the observed photometric data. We employ the Kurucz ATLAS9 stellar atmospheric models \citep{castelli1997} for the SED fitting, in which surface gravity (log~\textit{g}), [Fe/H] and T$_{eff}$ are the free parameters. The radius and mass can be estimated using the distance information. We have adopted a metallicity value of $[Fe/H] = -2.0$ for the cluster considering that the best-fit BaSTI isochrones provide a metallicity estimate of $[Fe/H] = -1.9$, and grid size in metallicity is 0.5 dex. Based on known values from literature, we have constrained the range of values of free parameters for the stellar populations. The values adopted and the results obtained are discussed below. \\

\begin{figure}
        \centering
        \includegraphics[width=\columnwidth]{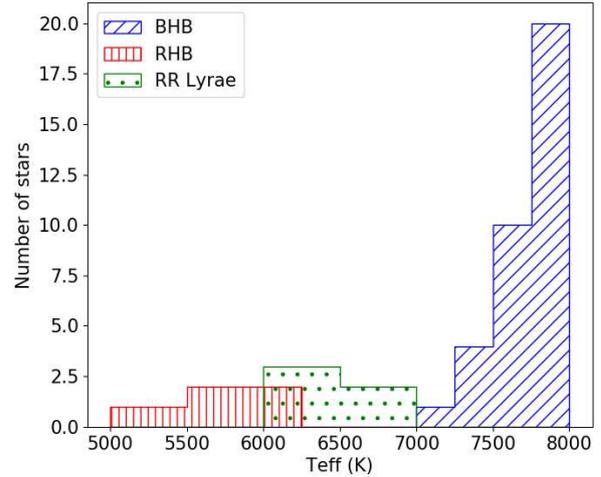}
        \caption{The histogram represents the distribution of effective temperature obtained from the best-fit models of UV-optical SEDs for the HB stellar population in NGC~5053.}
        \label{fig:fig8}
\end{figure}

\noindent \textit{Horizontal Branch stellar population}: BHB stars are known to have temperatures in the range $6000-20,000$~K and log~\textit{g} values between $3-5$ \citep{heber1997,kinmann2000,behr2003}. Values from literature suggest RHB stars to have effective temperatures in the range $5000-6000$~K and surface gravity values between 2 to 3 \citep{behr2003,sneden2010}. RR~Lyrae stars have been found to have intermediate values of parameters with their effective temperatures between $6000-8000$~K and surface gravity values between $2 -3$ \citep{behr2003_1}. The canonical mass estimates for the HB stars are $\sim 0.5-0.6$~M$_{\sun}$. Spectroscopic observations in several GCs suggest that for HB stars the masses range between $0.35-0.8$~M$_{\sun}$ with radii between $1.3-12.5$~R$_{\sun}$ \citep{heber1997}. Based on these values we constrain the free-parameter range of T$_{eff}$ to be $5000-20,000$~K and log~\textit{g} values in the range $2-5$. We then limit the models to those having masses between $0.5-0.8$~M$_{\sun}$ and obtain the best-fit model. From the best-fit SEDs, we find that the radii lie within $2.9-4$~R$_{\sun}$ for the BHB stars, within $5.8-7$~R$_{\sun}$ for the RR~Lyrae stars, and within $7.2-10.5$~R$_{\sun}$ for the RHB stars. We note that the values obtained are well within the predicted range. The surface gravity value for the BHB stars were found to be 3, while for RR~Lyrae stars and RHB stars, it was found to be 2.5. The effective temperature distribution from best-fit to SEDs provided a range of 5000~K to 8000~K for the HB stars. This distribution is shown in Fig.~\ref{fig:fig8}. 
 
\noindent EHB stars have their effective temperature in the range $20,000 < \textrm{T$_{eff}$} < 40,000$~K and the surface gravity values lie in the range $5-6$ \citep{2maxted2001,Moehler_2004,ostensen2009}. We follow a similar method of constraining the effective temperature to the observed range, and log~\textit{g} to 5. The best-fit SED model is characterised by an effective temperature of $23,000 \pm500$~K, $\mathrm{R} \sim 0.4$~R$_{\sun}$ and $M\sim 0.47 $~M$_{\sun}$.\\

\noindent \textit{The Blue Straggler Population:} From our UV CMDs, we have identified 8 BSS candidates. We use the photometric data to confirm the status of the BSS candidates by fitting their SEDs with models of BSS in VOSA. We have used the same input parameter range for the known BSS population as well as the BSS candidate population. The estimates for T$_{eff}$ for BSSs lie within a range of $6200 - 13400$~K, log~\textit{g} value lie within $2.6 - 5$, and their masses fall in the range $0.8 - 2$~M$_{\sun}$ corresponding to radii estimates of $1 - 4.7$~R$_{\sun}$ \citep{shara1997,2005Amarco,santucci2015,2016parada,2019raso}. For the best-fit SEDs, the known BSS stars have effective temperatures within 7750~K to 8250~K, and log \textit{g} within $3 - 4.5$. The estimates of mass and radius obtained are between $1.2-1.3$~M$_{\sun}$ and $0.9 - 2$~R$_{\sun}$, respectively. These estimates are consistent with the values discussed from literature and listed above. We affirm that the BSS candidates have their physical parameters within the range obtained for the known BSS stars in the cluster. We note that the residuals between the best-fit models and theoretical spectra are $\le15\%$. Thus, we confirm the status of BSS candidates from photometry as well as from the SED fitting of stellar atmosphere models. \\

\noindent \textit{AGB/PeAGB/e-BSS candidates:} There are 6 AGB/PeAGB/e-BSS candidates that we identified in the cluster. We observe these candidates to be cooler than the HB population and hotter than the giant population from the optical and UV CMDs. We, therefore, constrain the effective temperatures to a range of $3000-6000$~K, and an upper limit of surface gravity value to 4. On performing the SED fitting, we observe that these objects have surface gravity values between $2-3$. The effective temperature limits are within $4200 - 5400$~K for these objects. The mass estimates are at the higher end, between $1.55 - 2$~M$_{\sun}$, with radii values between $10 - 20$~R$_{\sun}$. We believe them to be e-BSSs, discussed later in Sect.~\ref{sec:disc}. 

\section{Discussion}
\label{sec:disc}

\par One of the major contributors to the UV emission from GCs is HB stars. The "first parameter" which determines the morphology of HB stars is metallicity \citep{1952arp}. Metal rich GCs are found to have a redder HB while metal poor GCs have bluer HB. This is because higher opacities in the metal rich environment makes the star appear cooler than the metal poor ones with the same envelope mass. But several authors have put forth observations of GCs with similar metallicity showing different HB morphologies. This is known as the "second parameter" effect \citep{1960Asandage}. Parameters like age, mass loss, initial He abundance, stellar rotation etc have been proposed as possible contributors to the second parameter effect \citep[see e.g.][]{2013dotter}. In our UV images, we have identified all the 35 BHB, 10 RR~Lyrae and 8 RHB stars. The spread of stars towards the bluer side of HB is larger and better populated compared to the red side. The blue dominated HB population is a typical characteristic of a metal poor GC. The HB sequence forms a horizontal sequence in the optical \textit{Gaia} CMD, while in the UVIT UV CMDs the HB sequence is more extended and delineates to a diagonal distribution. The BSS population occupies a parallel and diagonal sequence just below the HB in the UV CMDs. \citet{schiavon2012} had presented UV CMDs for galactic GCs using GALEX data. They had noted a similar observation of the HB stars forming a diagonal sequence and a parallel sequence of BSSs which are fainter than the HB.

\par
\begin{figure}
   \centering
    \includegraphics[width=\columnwidth]{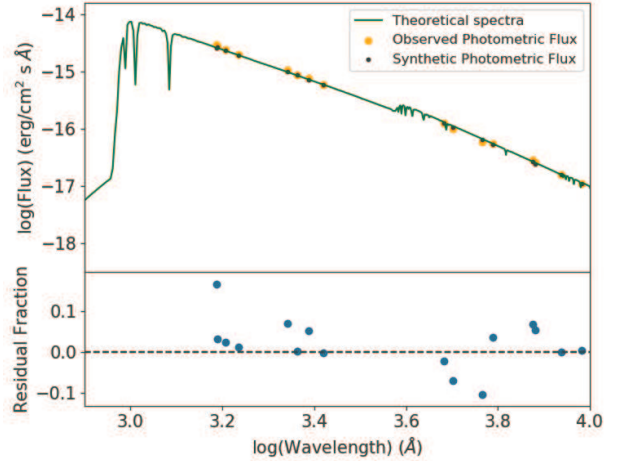}
   \caption{The spectral energy distribution (SED) of the EHB candidate constructed using GALEX, UVIT, Pan-STARRS and \textit{Gaia} is shown in the upper panel. The best-fit to the SED from Kurucz grid of models of stellar atmosphere is shown as solid line. The corresponding residuals are plotted in the lower panel.}
    \label{fig:fig9}
\end{figure}

 The location at the extreme hot end of the HB in both, the \textit{Gaia} optical CMD and UVIT UV CMDs, assisted us in identifying a candidate belonging to the EHB stellar population. EHB stars have He burning cores of masses close to He flash mass $\sim$0.47 M$_{\sun}$ with a very thin hydrogen envelope \citep{ostensen2009}. These stars have effective temperatures in the range $20,000 < T_{eff} < 40,000$~K and have been discovered in both metal poor and metal rich clusters \citep[e.g.][]{liebert1994,buo1986}. Though the evolution of EHB stars after He exhaustion in the core is well understood, a lot of uncertainty surrounds the origin of these stars. Broadly the formation mechanisms for EHB stars can be organized into two groups: (i) formation of single EHB stars \citep{1984webbink,1996dcruz,dcruz1996,1997sweigart,1998nelemans,soker98,saio2000,brown2001,2002antona}, (ii) formation of EHB stars in binary systems \citep{2003han,ostensen2009}. For the EHB candidate identified by us in NGC~5053, we generated the SED using photometric data from UVIT catalogue, \textit{Gaia} EDR3 catalogue, GALEX DR5 catalogue and Pan-STARRS DR1 catalogue as explained in Sect.~\ref{sec:stellarchar}. As discussed in Sect.~\ref{sec:stellarchar}, the EHB star is characterised by an $T_{eff} = 23,000$~K, log~\textit{g} = 5 and $[Fe/H] = -2.0$. For these parameters, a theoretical spectrum has been generated using the Kurucz ATLAS9 stellar atmospheric model and compared with the observed SED, shown in Fig.~\ref{fig:fig9}. We observe a good correlation between the model and the data by visualizing the residuals, which are found to be $\lesssim15\%$. This suggests that the EHB candidate detected in the current study is highly likely to be a single EHB star. We, therefore, believe the EHB star to have been formed in isolation.
\begin{figure*}
         \centering
         \includegraphics[width=\textwidth]{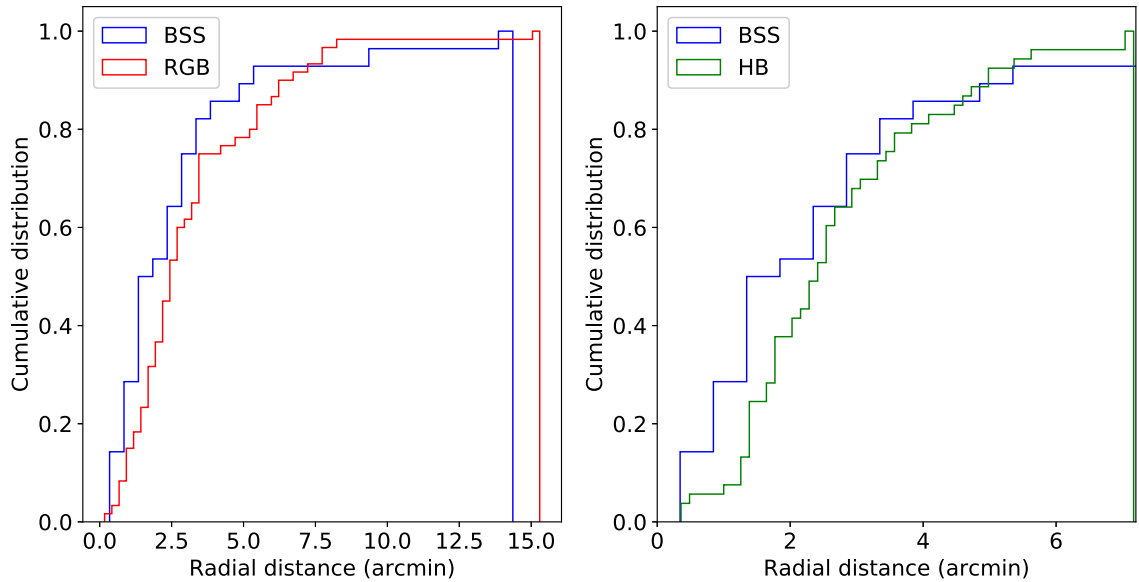}
         \caption{Cumulative radial distributions of the BSS population with respect to RGB (left) and HB (right) populations.}
         \label{fig:fig10}
\end{figure*}
\par Sixteen known BSSs have been located using the UVIT filters. In addition, eight have been identified as BSS candidates. The contention that they belong to the BSS stellar population in the cluster is supported with results from photometry as well as SED fitting.  
We analyse the distribution of the known BSSs and the BSS candidates in the cluster NGC~5053. We note that the known BSSs as well as most of the BSS candidates are concentrated within $6'$ from the center. Two BSS candidates are located further out, at radial distances of $9.5'$ and $14.4'$ from the centre. The statistical significance of the differences among the radial distribution of BSSs with respect to HB and RGB stars is assessed using the Kolmogorov-Smirnov (KS) test. The test reveals that the probability of BSSs and HB stars being drawn from the same population is negligible with a p-value~$\sim 7\times10^{-3}$. In the case where the RGB stars constitute the reference population, the KS test shows that the BSS population is significantly different from the RGB population with a p-value~$\sim 3\times10^{-4}$. The KS test thus demonstrates that the BSS population is significantly different from both RGB and HB population distributions. From the normalized cumulative radial distribution of the BSSs with respect to RGB and HB populations as shown in Fig.~\ref{fig:fig10}, it is evident that the BSSs are more centrally concentrated than the RGB and HB. The formation and evolution of the BSSs still remain open questions.
\par The formation scenarios of BSSs can be broadly classified into two categories: (i) a mass transfer (MT) scenario which is usually expected in a low density environment \citep{1964mccrea}, and (ii) a merger/collisional scenario which is expected in a dense environment \citep{1976hils}. Most studies agree that a combination of these scenarios is possible within a single GC. It is expected that the collisionally formed BSSs would be centrally concentrated as compared to those formed from MT mechanism. This is because there is a high probability of collisions to occur in the densely populated central regions of the cluster. The MT mechanism, on the other hand, will be relatively more efficient in dynamically relaxed regions \citep{2015davies}. In our case, we note that NGC~5053 is known to be a GC with a rather low central concentration, yet there is a high concentration of BSSs towards the centre. In addition to this, we also note that neither the BSSs nor the BSS candidates show a UV excess in their SEDs. An example of the SED for a star belonging to the BSS population of NGC~5053 is shown in Fig.~\ref{fig:fig11}. The other SEDs are available in electronic form, with a few sample SEDs in the appendix section (Fig.~\ref{appendix:sed}). This result would appear to suggest the absence of hot companions to these stars, thereby ruling out the MT formation mechanism \citep[e.g.][]{2014gosnell,2015gosnell,2016subra_astro}. We thus believe that the BSS population in NGC~5053 is plausibly formed as a result of collisions. This is also consistent with the speculation of \citet{leo1989} who had indicated a possibility of collisional origin for BSSs in low density clusters similar to the cases of NGC~5053 and NGC~5466. An alternate hypothesis is that primordial binaries could have given rise to these BSSs either through collisional and/or MT processes \citep{1991leofahl}.

\begin{figure}
   \centering
   \includegraphics[width=\columnwidth]{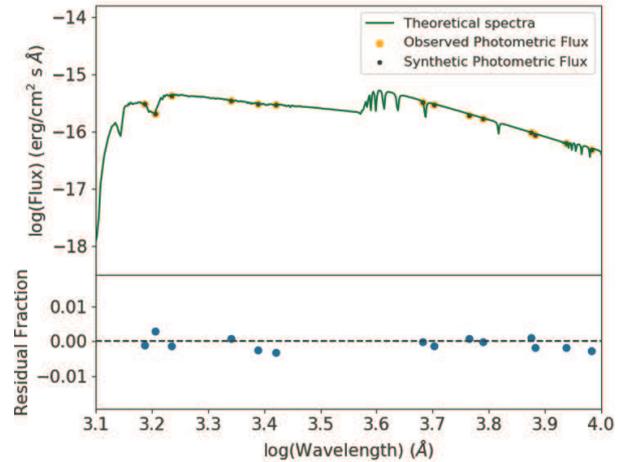}
   \caption{The spectral energy distribution (SED) of a BSS constructed using GALEX, UVIT, Pan-STARRS and \textit{Gaia} is shown in the upper panel. The best-fit to the SED from Kurucz grid of models of stellar atmosphere is shown as solid line. The corresponding residuals are plotted in the lower panel.}
    \label{fig:fig11}
\end{figure}
\par A group of unidentified stars lying above the RHB but hotter than the RGB in the optical \textit{Gaia} CMD was hypothesised as AGB/PeAGB/eBSS candidates. Our UV photometric results with UVIT filters show that these stars have FUV counterparts. This observation helps rule out their association to the AGB family as we do not expect the relatively cool AGB population to be bright at far UV wavelengths. The PeAGB stars have been found to to be much brighter than the HB in UV CMDs \citep{schiavon2012}, while the candidates in discussion are fainter and cooler than the HB. Thus their association to the PeAGB stellar population can be discarded. The only probable likelihood are of these stars to belong to the evolved phase of BSSs. Evidences of  existence of eBSS candidates in various clusters have been reported in literature. \citet{1988renzini} were the first to suggest that BSS stars in their core He burning phase will appear redder and brighter than the HB stars in the optical CMD. Their suggestion was based on the modeled magnitudes and colours of stars with masses up to twice the turn-off mass in GCs, which is typically the expected mass of BSSs. A more recent study by \citet{2009sills} produced similar results. eBSS candidates have been observed in clusters like M3, M80 and 47 Tuc \citep{1994bailyn,1999ferraro,2006beccari,2016ferraro}. The eBSSs are thought to lie in the region between sub-giant branch (SGB), RGB, BSS and HB in UV CMDs based on evolutionary models \citep{2016parada}. In our UV CMDs too, the eBSS candidates lie in the same region. We, therefore, believe this group of stars to be linked to the BSS progeny stellar population. This is consistent with the SED fitting results that indicate these stars to be more massive ($1.5-2$~M$_{\sun}$) and of larger size ($10-20$~M$_{\sun}$), suggesting that they belong to an evolved population. This is further corroborated by the fact that this cluster has a substantial BSS population. Follow-up spectroscopic studies are required to understand the true nature of these objects.

\section{Conclusion}
\label{sec:conc}
\par In this paper, we present the photometric analysis of the globular cluster NGC~5053 using the FUV and NUV filters of UVIT, in combination with \textit{Gaia} observations. We have characterized the UV bright population using UVIT, \textit{Gaia}, Pan-STARRS and GALEX data. Below is a summary of the important results from our work.
\begin{itemize}

    \item NGC~5053 has been observed in three FUV (F154W, F169M, F172M) and three NUV (N219M, N245M, N263M) filters of UVIT, on-board the \textit{AstroSat}. We constructed a catalogue of 1871 UV stars and carried out their photometry.
    
    \item Based on the PM of the stars from \textit{Gaia} EDR3 data, we identified the cluster members. We constructed UV CMDs and observed the cluster to have a significant blue HB which is typical of a metal poor GC. We also note that, unlike in optical CMDs, the HB sequence forms a broader and diagonal sequence in the UV CMDs with the BSS population lying parallel and fainter than the HB population sequence.
    
    \item On the basis of the location of the stars in the UV and optical CMDs we locate candidates of stellar populations that were not identified earlier in literature. They include (i)  8 BSS candidates, (ii) 6 eBSS candidates, and (iii) one EHB star. Further confirmation about their nature was obtained through fitting of SEDs using the Kurucz ATLAS9 stellar atmospheric models.
    
    \item The radial distribution and SEDs of the BSS population suggest a collisional scenario as the plausible mechanism of formation.
    
     \item Using an existing isochrone library, BaSTI-IAC, we characterized the cluster parameters. We found the best-fit model to be an $\alpha$ enhanced model of $[\alpha/Fe] = +0.4$, with He fraction = 0.247, $[Fe/H] = -1.9$, distance modulus $ = 16.16$ and age  $\sim12.5 \pm 2.0$~Gyr.
     
\end{itemize}

\section*{Acknowledgements}
\par We thank the referee for providing useful comments that
have helped in the improvement of the paper. This publication uses the data from the \textit{AstroSat} mission of the Indian Space Research Organisation (ISRO), archived at the Indian Space Science Data Centre (ISSDC). This publication uses UVIT data processed by the payload operations centre at IIA. The UVIT is built in collaboration between IIA, IUCAA, TIFR, ISRO and CSA.

\par This work has made use of BaSTI web tools. This publication makes use of VOSA, developed under the Spanish Virtual Observatory project supported by the Spanish MINECO through grant AyA2017-84089. In addition, this work has made use of data from the European Space Agency (ESA) mission {\textit Gaia} (\url{https://www.cosmos.esa.int/gaia}), processed by the {\textit Gaia} Data Processing and Analysis Consortium (DPAC, \url{https://www.cosmos.esa.int/web/gaia/dpac/consortium}). Funding for the DPAC has been provided by national institutions, in particular the institutions participating in the {\textit Gaia} Multilateral Agreement. This work has used observations made with the NASA \textit{Galaxy Evolution Explorer} (GALEX). GALEX is operated for NASA by the California Institute of Technology under NASA contract NAS5- 98034. The Pan-STARRS1 Surveys (PS1) and the PS1 public science archive have been made possible through contributions by the Institute for Astronomy, the University of Hawaii, the Pan-STARRS Project Office, the Max-Planck Society and its participating institutes, the Max Planck Institute for Astronomy, Heidelberg and the Max Planck Institute for Extraterrestrial Physics, Garching, The Johns Hopkins University, Durham University, the University of Edinburgh, the Queen's University Belfast, the Harvard-Smithsonian Center for Astrophysics, the Las Cumbres Observatory Global Telescope Network Incorporated, the National Central University of Taiwan, the Space Telescope Science Institute, the National Aeronautics and Space Administration under Grant No. NNX08AR22G issued through the Planetary Science Division of the NASA Science Mission Directorate, the National Science Foundation Grant No. AST–1238877, the University of Maryland, Eotvos Lorand University (ELTE), the Los Alamos National Laboratory, and the Gordon and Betty Moore Foundation. 

\section*{Data Availability}
The observational data underlying this article are available in the article and in its online supplementary material.


\bibliographystyle{mnras}
\bibliography{ngc5053_ref} 


\appendix

\section{Supplementary table and figures}
\begin{table*}
\centering
\begin{tabular}{lccccccccccc}
\hline
RA & DEC & N263M & N219M & N245M & F154W & F169M & F172M & $\Delta$ & Source\\
 (deg) & (deg) & (mag) & (mag) & (mag) & (mag) & (mag) & (mag) & (arcsec) &\\
\hline
        199.1605 & 17.6168 & 18.86 & 19.27 & 19.22 & 19.32 & 19.41 & 19.90 & 0.3 & BHB \\
        199.0744 & 17.6291 & 19.04 & 19.52 & 19.38 & 20.79 & 20.69 & 21.57 & 0.2 & BHB \\
        199.1006 & 17.6448 & 19.39 & 19.25 & 19.30 & 19.61 & 19.71 & 20.26 & 0.3 & BHB \\
        199.1579 & 17.6455 & 19.14 & 19.67 & 19.55 & 21.60 & 21.40 & - & 0.3 & BHB \\
        199.0885 & 17.6639 & 19.63 & 19.65 & 19.54 & 20.64 & 20.28 & 20.86 & 0.4 & BHB \\ \hline
        \hline
\end{tabular}
\caption{UVIT photometric catalogue of NGC~5053 : An example of the photometric catalogue of detected stars in the UVIT images of NGC~5053. The coordinates, UV magnitudes in the UVIT filters as well as the positional difference with respect to \textit{Gaia} counterparts ($\Delta$) are included in the catalogue. The last column indicates the nature of the source detected. The entire table is available online.}
\end{table*} \label{tab:photo_catalogue}

\begin{figure*}
\centering
\begin{subfigure}{0.3\textwidth}
    \includegraphics[width=\textwidth]{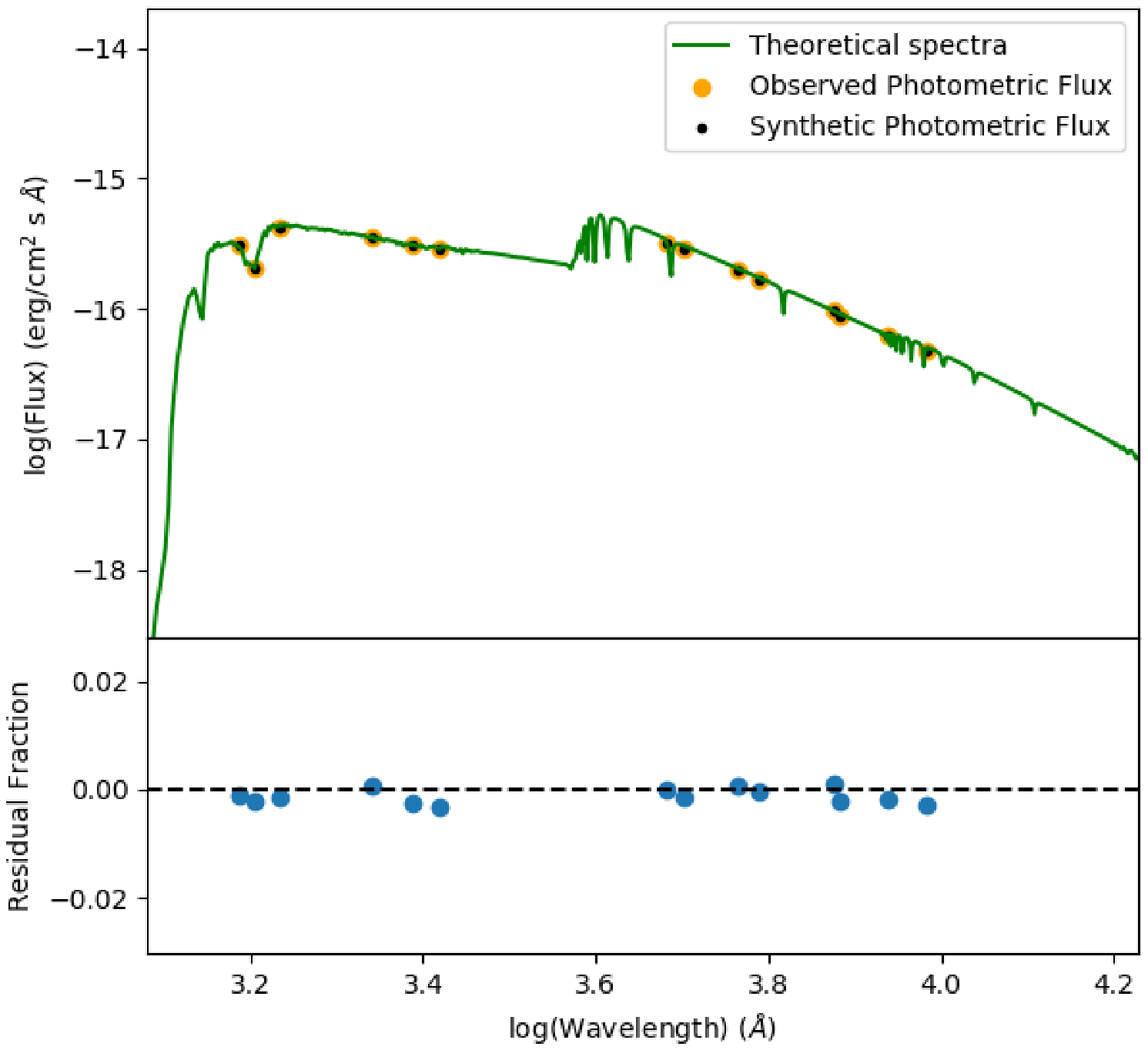}
    \caption{BSS 1}
\end{subfigure}
\hspace{-0.5cm}
\begin{subfigure}{0.3\textwidth}
    \includegraphics[width=\textwidth]{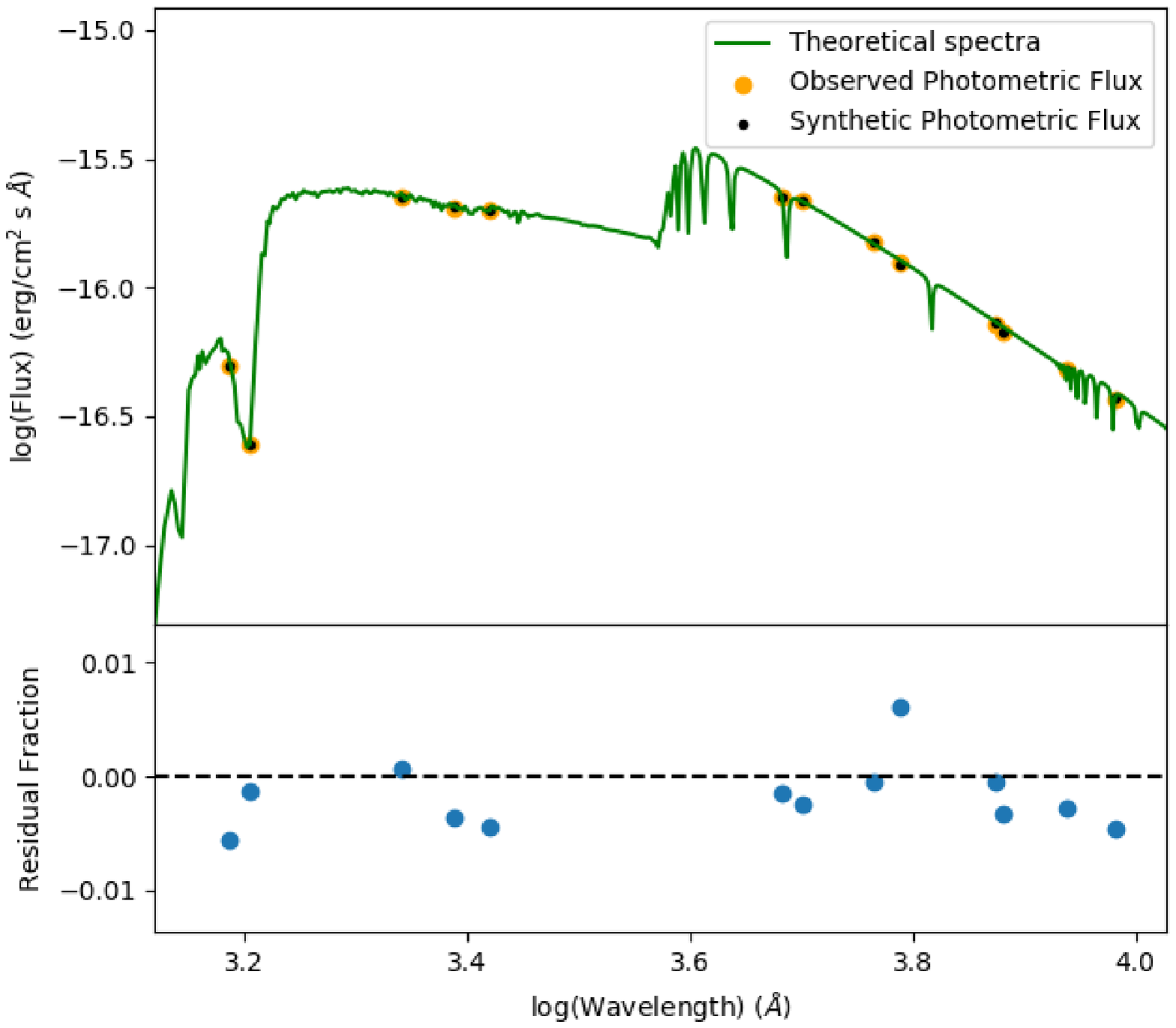}
    \caption{BSS 2}
\end{subfigure}
\hspace{-0.5cm}
\begin{subfigure}{0.3\textwidth}
    \includegraphics[width=\textwidth]{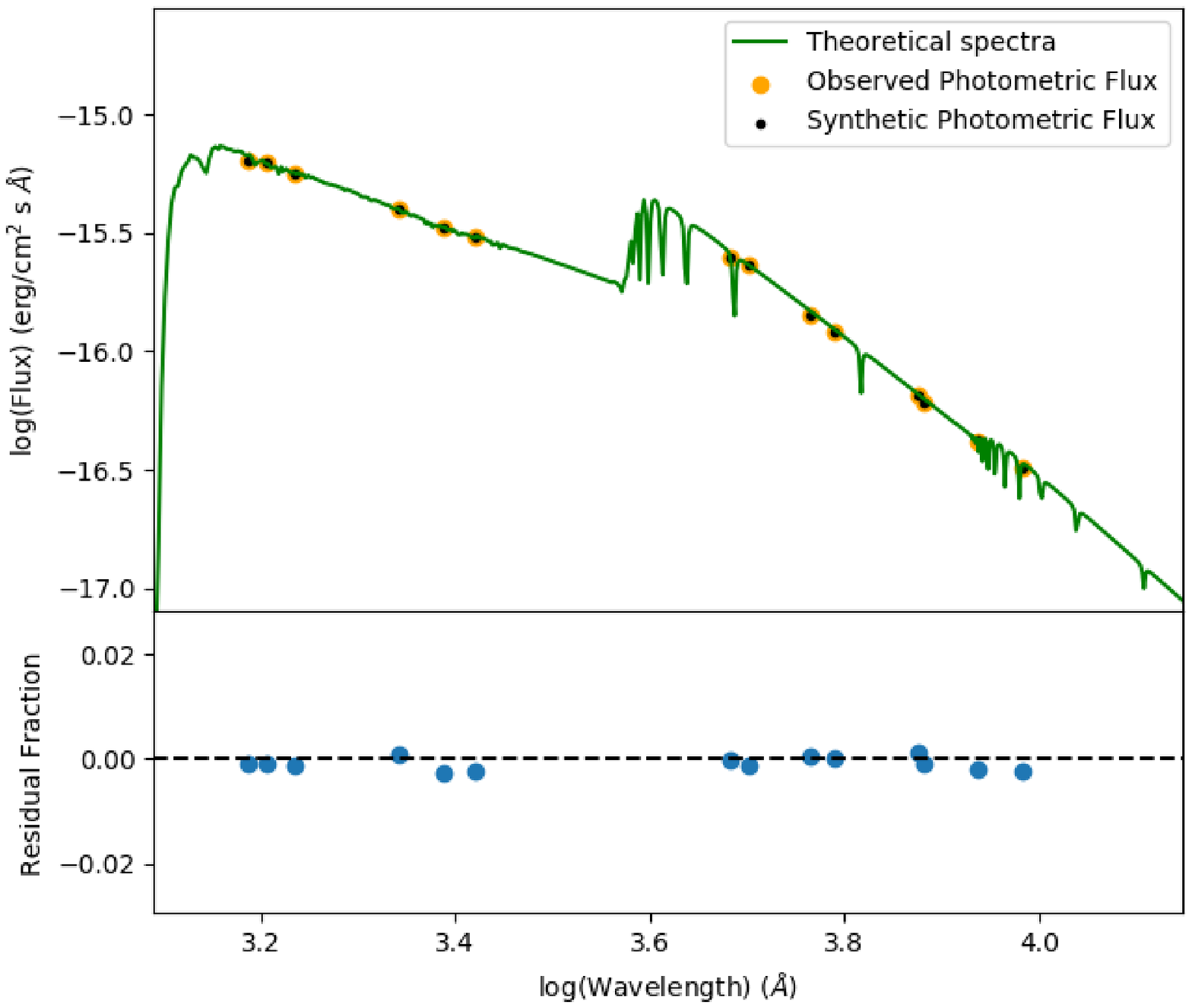}
    \caption{BSS 3}
\end{subfigure}
\hspace{-0.5cm}
\begin{subfigure}{0.3\textwidth}
    \includegraphics[width=\textwidth]{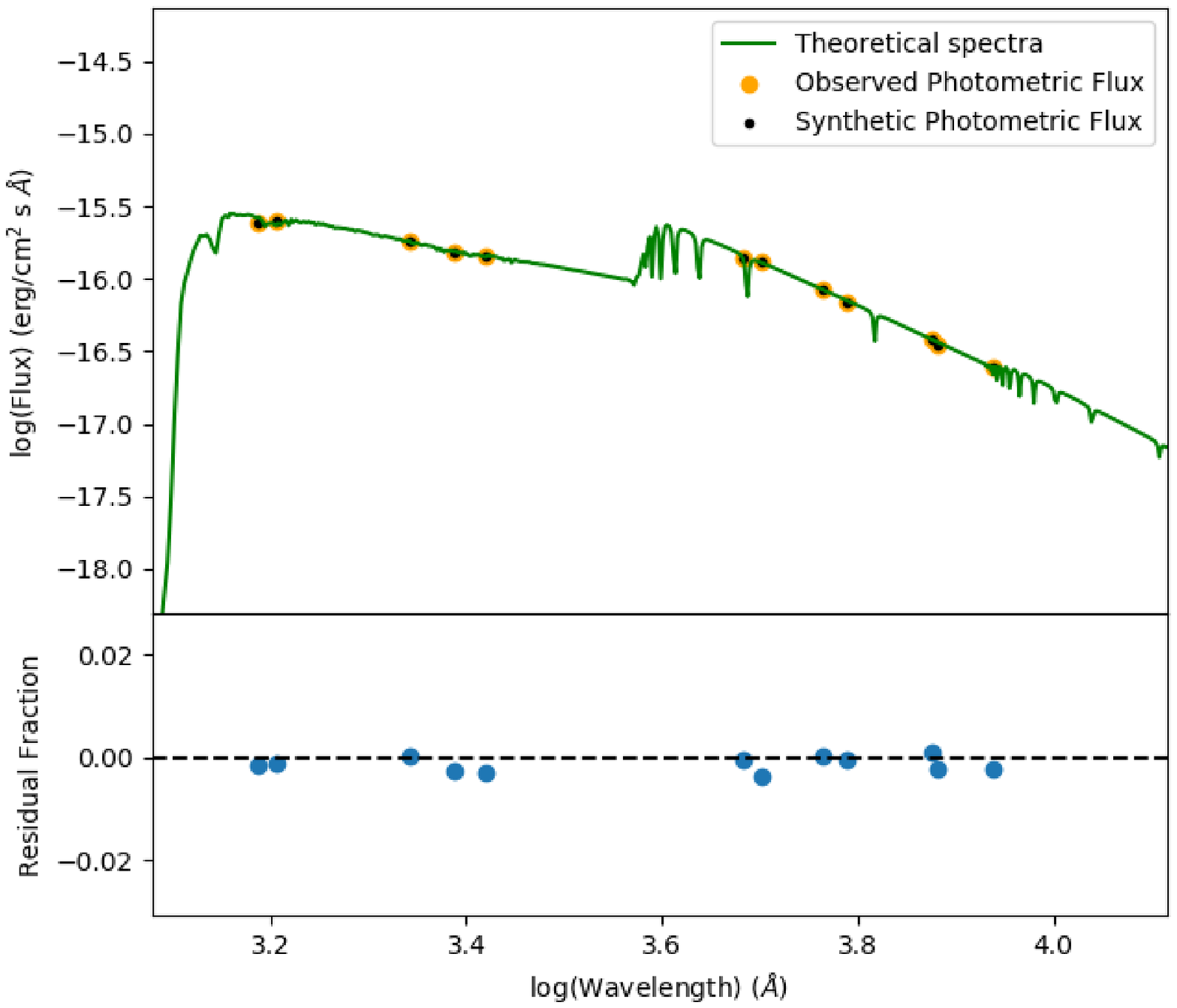}
    \caption{BSS 4}
\end{subfigure}
\hspace{-0.5cm}
\begin{subfigure}{0.3\textwidth}
    \includegraphics[width=\textwidth]{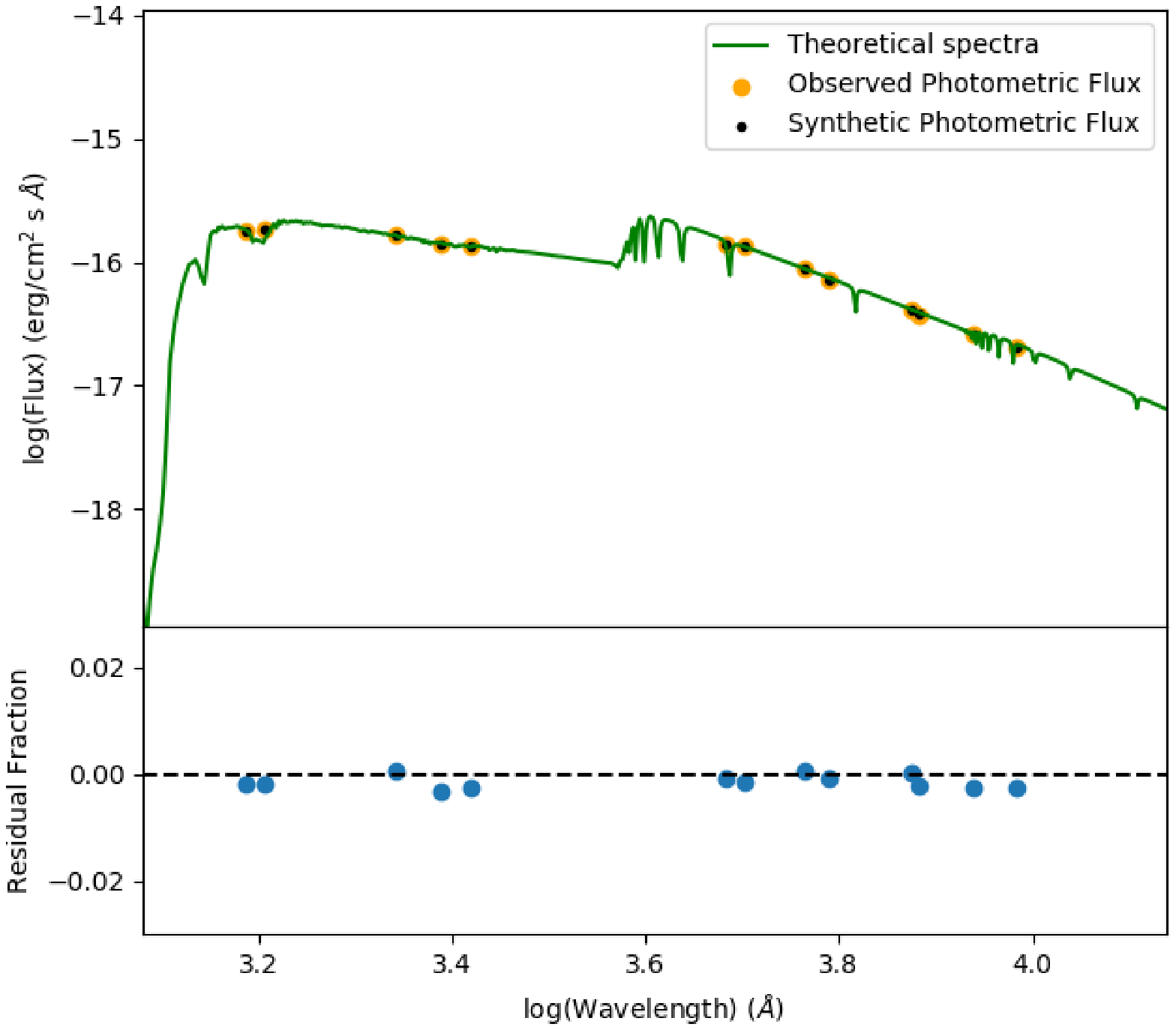}
    \caption{BSS 5}
\end{subfigure}
\hspace{-0.5cm}
\begin{subfigure}{0.3\textwidth}
    \includegraphics[width=\textwidth]{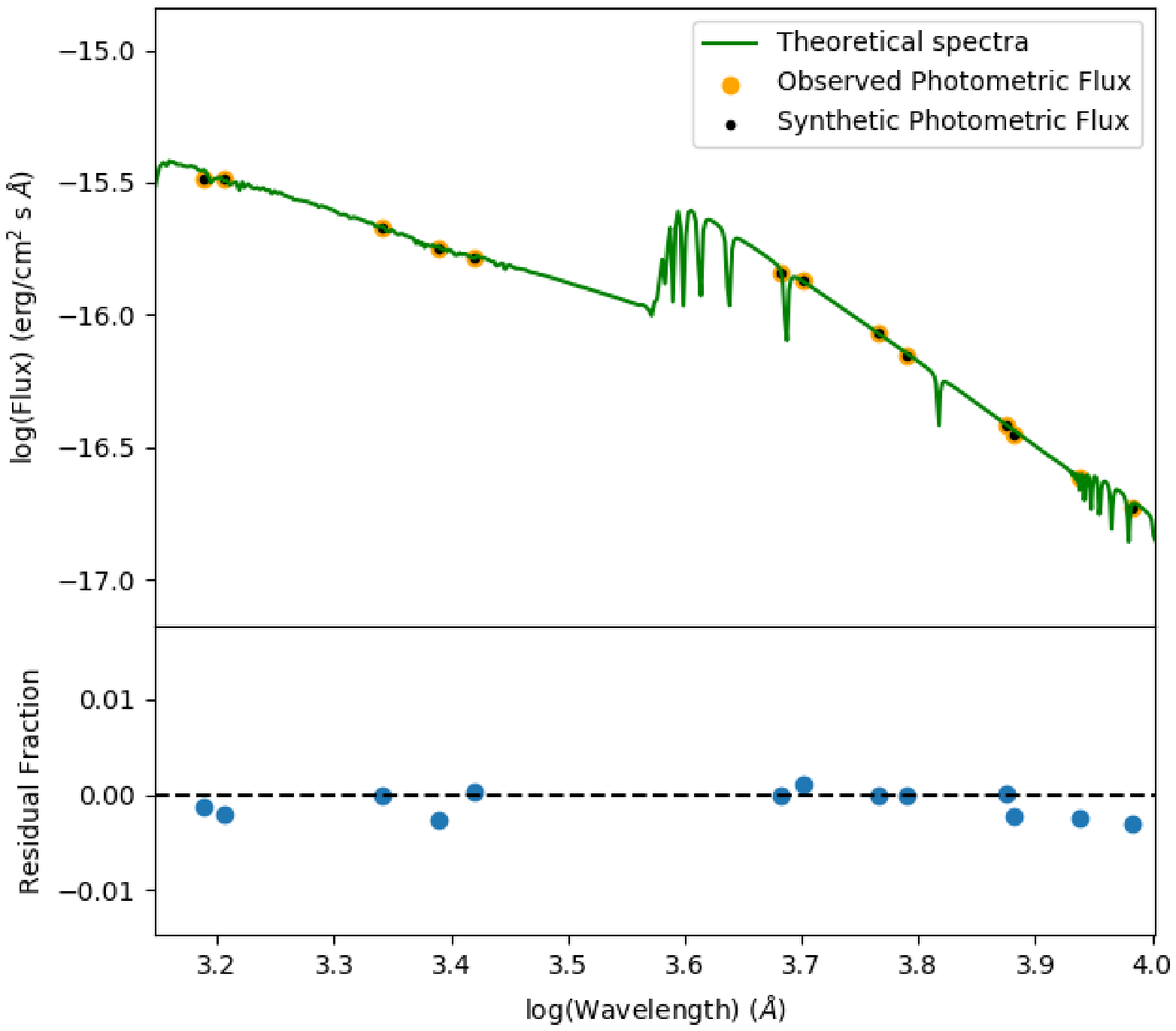}
    \caption{BSS 6}
\end{subfigure}
\hspace{-0.5cm}
\caption{Spectral energy distributions of few BSSs in NGC~5053. The entire figure is available online.}
\label{appendix:sed}
\end{figure*}


\bsp	
\label{lastpage}
\end{document}